\documentclass[sigplan,screen]{acmart}

\copyrightyear{2023} 
\acmYear{2023} 
\setcopyright{rightsretained} 
\acmConference[PPoPP '23]{The 28th ACM SIGPLAN Annual Symposium on Principles and Practice of Parallel Programming}{February 25-March 1, 2023}{Montreal, QC, Canada}
\acmBooktitle{The 28th ACM SIGPLAN Annual Symposium on Principles and Practice of Parallel Programming (PPoPP '23), February 25-March 1, 2023, Montreal, QC, Canada}
\acmDOI{10.1145/3572848.3577391}
\acmISBN{979-8-4007-0015-6/23/02}

\usepackage{booktabs}   
\usepackage{subcaption} 
\usepackage{algorithmic}
\usepackage{graphicx}
\usepackage{textcomp}
\usepackage{xcolor}
\usepackage{multirow}
\usepackage{amsthm}
\usepackage{array}
\usepackage[T1]{fontenc}
\usepackage{makecell}

\newcolumntype{P}[1]{>{\centering\arraybackslash}p{#1}}

\def\BibTeX{{\rm B\kern-.05em{\sc i\kern-.025em b}\kern-.08em
		T\kern-.1667em\lower.7ex\hbox{E}\kern-.125emX}}
	
\begin{document}

\title[W\textsc{rht}]{POSTER: Efficient All-reduce for Distributed DNN Training in Optical Interconnect Systems}         

\settopmatter{authorsperrow=1} 
\newcommand{\tsc}[1]{\textsuperscript{#1}} 
\author{Fei Dai\tsc{1}*, Yawen Chen\tsc{1}, Zhiyi Huang\tsc{1}, Haibo Zhang\tsc{1}, Fangfang Zhang\tsc{2}}
\affiliation{
	\institution{\tsc{1}University of Otago, Dunedin, New Zealand \tsc{2}Qilu University of Technology, Jinan, China}
	\city{}
	\country{} 
}

\email{daitr616@student.otago.ac.nz*,  {yawen.chen, zhiyi.huang, haibo.zhang}@otago.ac.nz,  zhff4u@qlu.edu.cn} 

\begin{abstract}
All-reduce is the crucial communication primitive to reduce model parameters in distributed Deep Neural Networks (DNN) training. Most existing all-reduce algorithms are designed for traditional electrical interconnect systems, which cannot meet the communication requirements for distributed training of large DNNs due to the low data bandwidth of the electrical interconnect systems. One of the promising alternatives for electrical interconnect is optical interconnect, which can provide high bandwidth, low transmission delay, and low power cost. We propose an efficient scheme called W\textsc{rht} (Wavelength Reused Hierarchical Tree) for implementing all-reduce operation in optical interconnect systems. W\textsc{rht} can take advantage of WDM (Wavelength Division Multiplexing) to reduce the communication time of distributed data-parallel DNN training.  Simulations using real DNN models show that, compared to all-reduce algorithms in the electrical and optical network systems, our approach reduces communication time by 75.76\% and 91.86\%, respectively.
\end{abstract}




\maketitle

\vspace{-1mm}
\section{Introduction}
\noindent Data parallelism is one of the most widely adopted paradigms where each worker trains the DNN using its local dataset and exchanges model parameters (e.g., gradients) with other workers iteratively~\cite{zhang2021near}. Stochastic Gradient Descent (SGD), the most widespread method for DNN training, intensively invokes data communications for all-reduce operations in distributed deep learning (DL)~\cite{huang2021communication}.
All-reduce aims to make every worker receive the model parameters from all the other workers and then apply the reduction operation to get the averaged model parameters. 
It has been shown that the communications for all-reduce with a large number of workers may occupy 50-90\% of per-iteration training time in current traditional electrical networks~\cite{wang2019blink}.
The communication time in traditional electrical interconnect can be severely high due to the low bandwidth of electrical routers, high latency of electrical networking, and network congestion. When the overhead caused by communication exceeds the gain brought by the parallel computation, the training performance will be degraded. 
With the recent development of CMOS-compatible optical devices~\cite{yang2019multidomain}, optical intra/inter-chip network connection is a promising alternative, which can provide high bandwidth, low transmission delay, and low power cost. Moreover, optical interconnect can transmit data through a waveguide using different wavelengths enabled by leveraging WDM, enabling parallel data transmission. 

However, most existing all-reduce algorithms are not designed for optical interconnects. They are designed for electrical interconnect systems and do not take advantage of optical features such as parallel data transmission with WDM.
For instance, the well-known Ring all-reduce algorithm takes $2(n-1)$ steps to finish the all-reduce communications~\cite{ring}, where $n$ is the number of workers. However, such a method is unsuitable for optical interconnect systems because it only assumes one wavelength for transmission in each step, failing to take advantage of the WDM of optical interconnect. Therefore, we propose an efficient all-reduce scheme named W\textsc{rht} in an optical ring interconnect system with the objective of minimizing the number of communication steps and communication time for the all-reduce operation. As far as we know, W\textsc{rht} is the first scheme for optimizing all-reduce in optical interconnect systems.

\vspace{-3mm}
\section{ The W\textsc{rht} Scheme}\label{sec:performance} 
\begin{figure}[!hpbt]
	\vspace{-1mm}
	\includegraphics[scale=0.53]{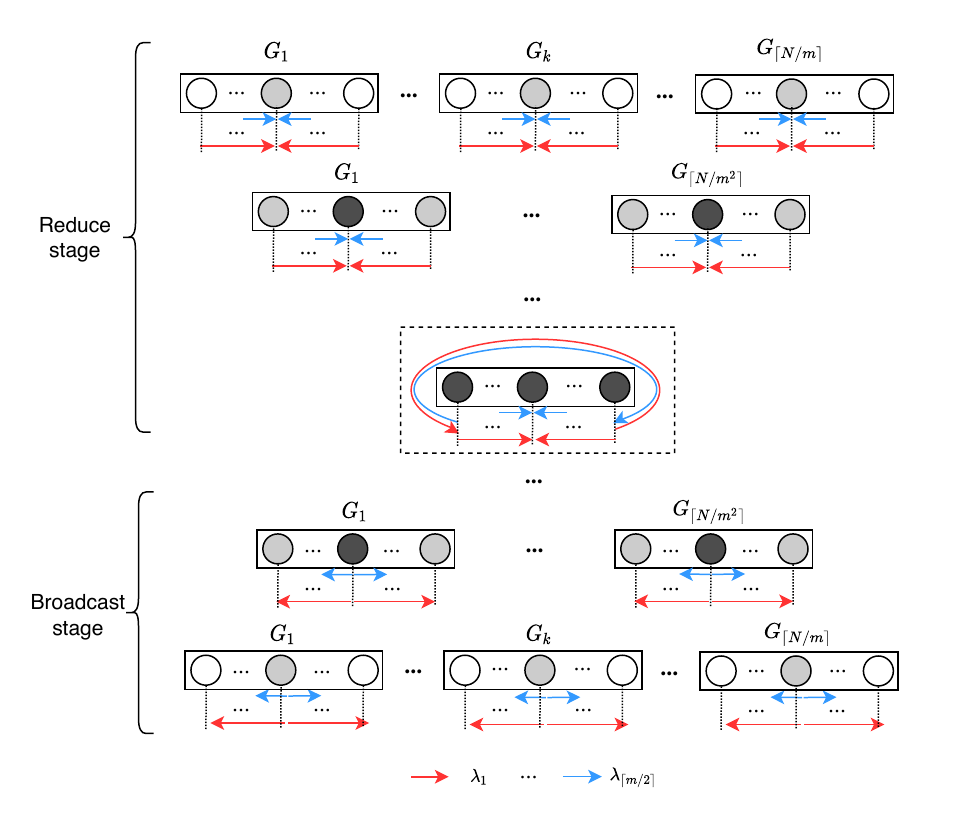}
	{
		\vspace{-7mm}
		\caption{The working principle in W\textsc{rht} scheme}
		\label{fig:WRHT}
	}
	\vspace{-2mm}
\end{figure}

\noindent W\textsc{rht} scheme is based on micro-ring resonator optical interconnect architecture called TeraRack~\cite{terarack}. We assume $N$ computing nodes are connecting with each other sequentially into a ring, and the computing node is GPU. The number of available wavelengths per waveguide is $w$, and the bandwidth per wavelength is $B$. 
We use Figure~\ref{fig:WRHT} to illustrate the mechanism of W\textsc{rht}, which consists of two stages: reduce
stage and broadcast stage. \\
\textbf{Reduce stage:}
In step 1, all nodes are partitioned into groups along the ring with each group having $m$ nodes. 
The intermediate node of each group is selected as the representative node and responsible for collecting the data within each group by $\lfloor m/2 \rfloor$ wavelengths. After that, each representative node executes a reduction operation to be transmitted in the next step. 
In the subsequent step $i$, the old representative nodes selected in the previous step are further partitioned into $\lceil \frac{N}{m^{i}} \rceil$ groups and the middle node of each group is selected as the new representative node as illustrated in Figure~\ref{fig:WRHT}. 
This process is repeated until the wavelength is sufficient enough to provide all-to-all communication among the representative nodes in the last step, as illustrated by the dotted box in the middle of Figure~\ref{fig:WRHT}. 
\\
\textbf{Broadcast stage:}
Once the representative node(s) in the final step of reduce stage obtain the final reduction value, the process of broadcast stage is the reverse of reduce stage. Specifically, the representative nodes broadcast the reduction data in corresponding groups using $\lfloor m/2 \rfloor$ wavelengths, which is repeated according to the hierarchical tree structure until all the nodes receive the reduce data, as illustrated in the lower part of Figure 3. As a result, the total number of communication steps for W\textsc{rht} is $2\lceil \log_{m}N \rceil$ or $2\lceil \log_{m}N \rceil - 1$.

Since nodes are partitioned into subgroups in each step by W\textsc{rht}, communications within each subgroup need to be assigned with proper wavelengths to avoid conflicts, while communications from different subgroups do not share any link between groups on the ring topology. Therefore, we can use wavelength assignment approach First Fit~\cite{ozdaglar2003routing} or Best Fit~\cite{sathishkumar2015best} for assigning the wavelengths within each subgroup. As the number of nodes in each subgroup is $m$ and the intermediate node is selected as the representative node, it is easy to derive that the wavelength requirement is $\lfloor m/2 \rfloor$. For the last step in reduce stage, the number of representative nodes can be derived as $m^*=\lceil \frac{N}{m^{\lceil log_{m}N \rceil-1}} \rceil$, which requires $\lceil\frac{(m^*)^2}{8}\rceil $ wavelengths for all-to-all communications~\cite{liang2006general} when $m^*>1$.

\section{Experimental Setup and Results}   \label{sec:evaluation} 
DNN models used in the simulation are AlexNet (62.3M parameters)~\cite{krizhevsky2012imagenet}, VGG16 (138M parameters)~\cite{simonyan2014very}, ResNet50 (25M parameters)~\cite{he2016deep} and GoogLeNet (6.7977M parameters)~\cite{szegedy2015going} with ImageNet dataset~\cite{deng2009imagenet}. 
We implement W\textsc{rht} along with a list of all-reduce algorithms in our optical interconnect simulator, and we use SimGrid ~\cite{casanova2008simgrid} to simulate the electrical network system. We estimate the communication time by numerically setting different transferred data of DNNs, number of nodes, wavelengths, etc. in our simulator and SimGrid.

\begin{figure}[ht]
	\vspace{-2mm}
	\centering
	\includegraphics[scale=0.38]{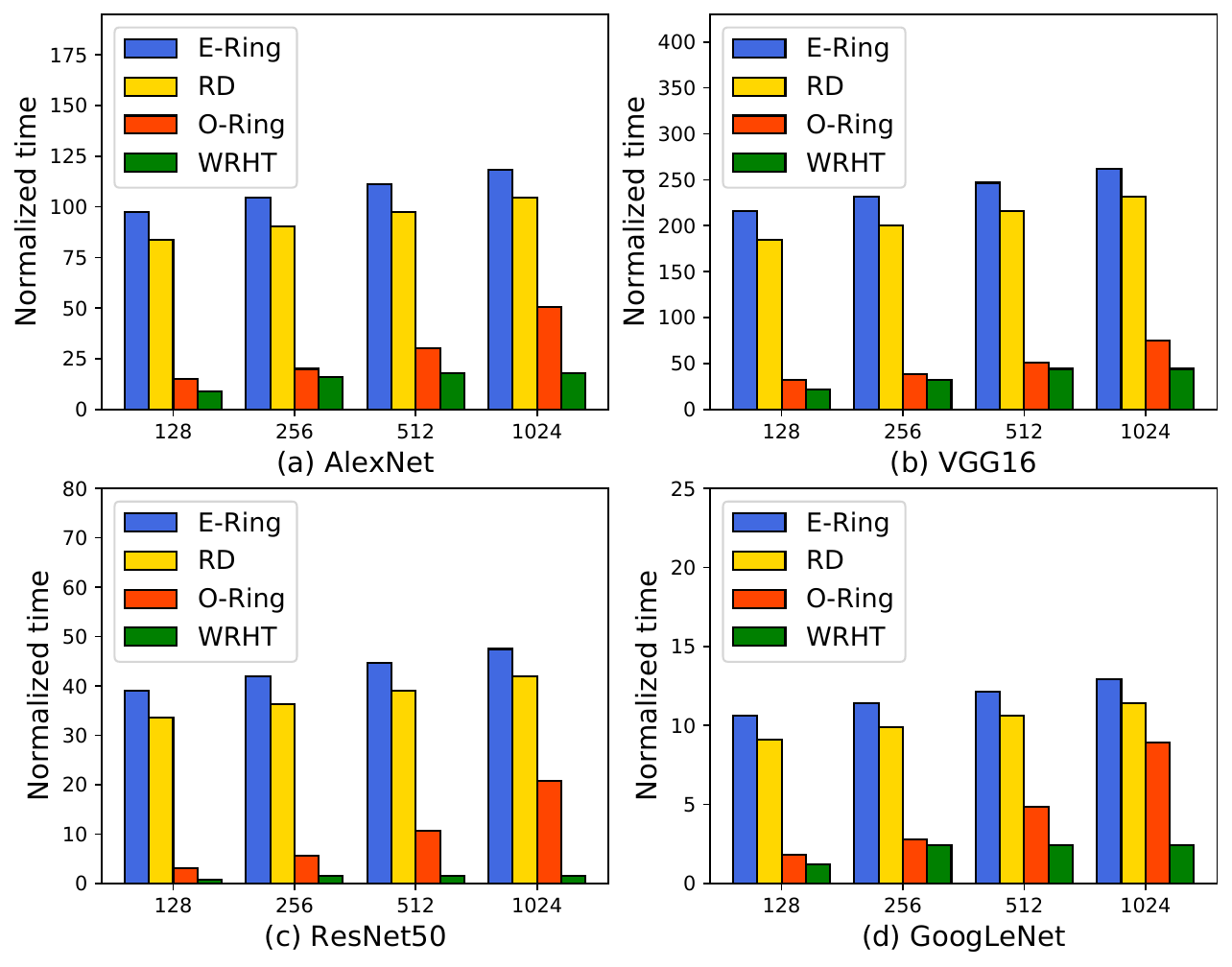}
	{
		\vspace{-2mm}
		\caption{The comparison of communication time in electrical interconnect and optical interconnect system using different all-reduce algorithms}
		\label{fig:comparison}
	}
\end{figure}

Figure~\ref{fig:comparison} compares the communication time of Ring and RD all-reduce algorithms in electrical interconnect system with Ring all-reduce and W\textsc{rht} in the optical interconnect system by different DNN models across different scales.

\vspace{-1mm}
\section{Conclusion}
In this paper, we propose an efficient all-reduce algorithm in optical interconnect systems called W\textsc{rht} by taking advantage of the multiple wavelengths to reduce the total number of communication steps. 
We have shown that our method significantly outperforms the all-reduce algorithms in electrical and optical interconnect systems with 75.76\% and 91.86\% communication time reduction. 


\bibliographystyle{unsrt}
\bibliography{ppopp_2}

\begin{thebibliography}{10}

\bibitem{zhang2021near}
Zhe Zhang, Chuan Wu, and Zongpeng Li.
\newblock Near-optimal topology-adaptive parameter synchronization in
  distributed dnn training.
\newblock In {\em IEEE INFOCOM 2021-IEEE Conference on Computer
  Communications}, pages 1--10. IEEE, 2021.

\bibitem{huang2021communication}
Jiayi Huang, Pritam Majumder, Sungkeun Kim, Abdullah Muzahid, Ki~Hwan Yum, and
  Eun~Jung Kim.
\newblock Communication algorithm-architecture co-design for distributed deep
  learning.
\newblock In {\em 2021 ACM/IEEE 48th Annual International Symposium on Computer
  Architecture (ISCA)}, pages 181--194. IEEE, 2021.

\bibitem{wang2019blink}
Guanhua Wang, Shivaram Venkataraman, Amar Phanishayee, Jorgen Thelin, Nikhil
  Devanur, and Ion Stoica.
\newblock Blink: Fast and generic collectives for distributed ml.
\newblock {\em arXiv preprint arXiv:1910.04940}, 2019.

\bibitem{yang2019multidomain}
Peng Yang, Zhehui Wang, Zhifei Wang, Jiang Xu, Yi-Shing Chang, Xuanqi Chen,
  Rafael~KV Maeda, and Jun Feng.
\newblock Multidomain inter/intrachip silicon photonic networks for
  energy-efficient rack-scale computing systems.
\newblock {\em IEEE Transactions on Computer-Aided Design of Integrated
  Circuits and Systems}, 39(3):626--639, 2019.

\bibitem{ring}
Pitch Patarasuk and Xin Yuan.
\newblock Bandwidth optimal all-reduce algorithms for clusters of workstations.
\newblock {\em Journal of Parallel and Distributed Computing}, 69(2):117--124,
  2009.

\bibitem{terarack}
Mehrdad Khani, Manya Ghobadi, Mohammad Alizadeh, Ziyi Zhu, Madeleine Glick,
  Keren Bergman, Amin Vahdat, Benjamin Klenk, and Eiman Ebrahimi.
\newblock Terarack: A tbps rack for machine learning training.
\newblock 2020.

\bibitem{ozdaglar2003routing}
Asuman~E Ozdaglar and Dimitri~P Bertsekas.
\newblock Routing and wavelength assignment in optical networks.
\newblock {\em IEEE/ACM transactions on networking}, 11(2):259--272, 2003.

\bibitem{sathishkumar2015best}
P~Sathishkumar and V~Mahalingam.
\newblock Best-fit wavelength assignment algorithm for persistent communication
  in optical networks.
\newblock {\em International Journal of Computer Science and Information
  Technologies}, 6(1):728--733, 2015.

\bibitem{liang2006general}
Weifa Liang and Xiaojun Shen.
\newblock A general approach for all-to-all routing in multihop wdm optical
  networks.
\newblock {\em IEEE/ACM transactions on networking}, 14(4):914--923, 2006.

\bibitem{krizhevsky2012imagenet}
Alex Krizhevsky, Ilya Sutskever, and Geoffrey~E Hinton.
\newblock Imagenet classification with deep convolutional neural networks.
\newblock {\em Advances in neural information processing systems},
  25:1097--1105, 2012.

\bibitem{simonyan2014very}
Karen Simonyan and Andrew Zisserman.
\newblock Very deep convolutional networks for large-scale image recognition.
\newblock {\em arXiv preprint arXiv:1409.1556}, 2014.

\bibitem{he2016deep}
Kaiming He, Xiangyu Zhang, Shaoqing Ren, and Jian Sun.
\newblock Deep residual learning for image recognition.
\newblock In {\em Proceedings of the IEEE conference on computer vision and
  pattern recognition}, pages 770--778, 2016.

\bibitem{szegedy2015going}
Christian Szegedy, Wei Liu, Yangqing Jia, Pierre Sermanet, Scott Reed, Dragomir
  Anguelov, Dumitru Erhan, Vincent Vanhoucke, and Andrew Rabinovich.
\newblock Going deeper with convolutions.
\newblock In {\em Proceedings of the IEEE conference on computer vision and
  pattern recognition}, pages 1--9, 2015.

\bibitem{deng2009imagenet}
Jia Deng, Wei Dong, Richard Socher, Li-Jia Li, Kai Li, and Li~Fei-Fei.
\newblock Imagenet: A large-scale hierarchical image database.
\newblock In {\em 2009 IEEE conference on computer vision and pattern
  recognition}, pages 248--255. Ieee, 2009.

\bibitem{casanova2008simgrid}
Henri Casanova, Arnaud Legrand, and Martin Quinson.
\newblock Simgrid: A generic framework for large-scale distributed experiments.
\newblock In {\em Tenth International Conference on Computer Modeling and
  Simulation (uksim 2008)}, pages 126--131. IEEE, 2008.

\end{thebibliography}

%

\end{document}